\documentclass[prl,amsmath,showpacs,superscriptaddress]{revtex4}
\usepackage{amsmath}
\usepackage{amsbsy}
\usepackage{amscd}
\usepackage{amsopn}
\usepackage{amstext}
\usepackage{amsxtra}
\usepackage{graphicx}
\usepackage{epsfig}
 \usepackage{hyperref}
 \newtheorem{thm}{Lemma}

\begin{document}
\title{Competition-induced criticality in a model of meme popularity}
\author{James P.~Gleeson}
\affiliation{MACSI, Department of Mathematics \& Statistics, University of Limerick, Ireland}
\author{Jonathan A.~Ward}
\affiliation{Centre for the Mathematics of Human Behaviour, Department of Mathematics and
Statistics, University of Reading, Whiteknights, UK}
\author{Kevin~P.~O'Sullivan}
\affiliation{MACSI, Department of Mathematics \& Statistics, University of Limerick, Ireland}
\author{William T.~Lee}
\affiliation{MACSI, Department of Mathematics \& Statistics, University of Limerick, Ireland}
\date{24 Dec 2013}%

% 64.60.aq}
\begin{abstract}
Heavy-tailed distributions of meme popularity occur naturally in a model of meme diffusion on social networks. Competition between multiple memes for the limited resource of user attention is identified as the mechanism that poises the system at criticality. The popularity growth of each meme is described by a critical branching process, and asymptotic analysis predicts  power-law distributions of popularity with very heavy tails (exponent $\alpha<2$, unlike preferential-attachment models), similar to those seen in empirical data.
\end{abstract}
\maketitle

When people select from multiple items of roughly equal value, some items quickly become  extremely popular, while other items are chosen by relatively few people \cite{Bentleybook}.
%The probability $P(n)$ that a random item has popularity $n$ at time $t$---where the popularity of an item is the number of times it was selected prior to time $t$---is often observed to have a heavy-tailed distribution. % cf Porter
The probability $P_n(t)$ that a random item has been selected $n$ times by time $t$ is often observed to have a heavy-tailed distribution ($n$ is called the \emph{popularity} of the item at time $t$).
In examples where the items are baby names \cite{Bentley04}, apps on Facebook \cite{Onnela10}, retweeted URLs or hashtags on Twitter \cite{Lerman12,Bakshy11,Banos13}, or video views on YouTube \cite{Szabo10}, the popularity distribution  is found to scale approximately as a power-law $P_n \sim n^{-\alpha}$ over several decades. The exponent $\alpha$ in all these examples is less than 2, and typically has a value close to 1.5.
This range of $\alpha$ values is notably distinct from those obtainable from cumulative-advantage or preferential-attachment models of Yule-Simon type---as used to describe power-law degree distributions of networks, for example \cite{Redner98,Newman05,Simkin11,Barabasi99}---which give $\alpha\ge2$. Interestingly, the value $\alpha=1.5$ is also found for the power-law distribution of avalanche sizes in self-organized criticality (SOC) models \cite{Zapperi95,Bakbook},  suggesting the possibility that the heavy-tailed distributions of popularity in the examples above are due to the systems being somehow poised at criticality.

In this paper we present an analytically tractable model of selection behaviour, based on simplifying the model of Weng et.~al \cite{Weng12} for the spreading of memes  on a social network. We show that in certain limits the system is automatically poised at criticality---in the sense that meme popularities are described by a critical branching process \cite{Harrisbook}---and that the criticality can be ascribed to the competition between memes for the limited resource of user attention. We dub this mechanism \emph{competition-induced criticality} (CIC) and investigate the impact of the social network topology (degree distribution) and the age of the memes upon the distribution of meme popularities. { We show that  CIC  gives rise to heavy-tailed distributions very similar to the distributions of avalanche sizes in SOC models \cite{Goh03,Noel13}, even though our competition mechanism is quite different from the sandpile paradigm of SOC. This work may therefore be of interest in other areas where SOC-like critical phenomena have been observed in experiments or simulations, such as economic models of competing firms \cite{Krider97,Arenas02}, the evolution and extinction of competing species \cite{Sprott04,Newman96,Sole96}, and neural activity in the brain \cite{Haimovici13,Rubinov11}.
}

\begin{figure}
\centering
\epsfig{figure=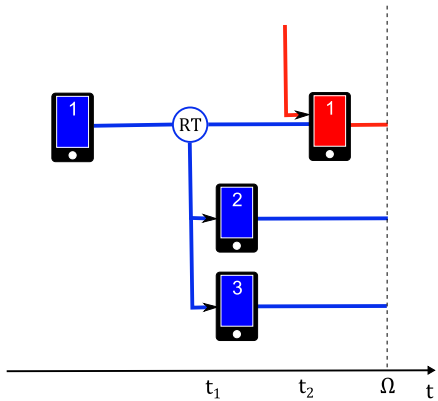,width=5.7cm}
\caption{{ Schematic of the model. Time runs horizontally and nodes of the network are listed vertically; the screen colour of each node indicates the meme it currently holds. At time $t_1$, node 1 retweets the blue meme to its followers (nodes 2 and 3). At time $t_2$, node 1's screen is overwritten by the red meme, which was tweeted by one of the nodes followed by node 1.}
%At the observation time $\Omega$, the data on meme retweets for all earlier times is gathered and the popularities of all memes are compared.
}\label{figfromS1}
\end{figure}
For clarity, we will phrase the model in terms of meme diffusion as in \cite{Weng12} but the same understanding of the basic mechanism---and the analytical techniques for time-dependent distributions---can also be applied to other  models, including the random-copying popularity models of \cite{Bentley04,Bentley11,Evans07}. The role of competition among items for limited resources has been examined from many viewpoints: see, for example, \cite{Adami02,beguerisse10,Forgerini11} and also related work on competing diseases \cite{Karrer11,Noel12,Miller12}.
The distribution of popularity increments (number of selections of an item in a small time interval) in Moran-type models has been obtained analytically \cite{Evans07}; however, our focus is on the (time-dependent) distributions of popularity accumulated over  long timescales.
%---we believe this is the first model where time-dependent popularity distributions can be obtained using an analytical approach.

We consider a model of a directed social network, like Twitter, where nodes represent users; there are $N$ nodes and we will take the limit $N\to\infty$ in our analysis. A randomly-chosen user has $k$ \emph{followers} (i.e., out-degree $k$, note we use the convention that network edges are directed from nodes to their followers) with probability $p_k$.
%; the $k$ followers are chosen at random from the other $N-1$ nodes, giving a Poisson distribution of in-degrees.
 %The mean out-degree $z$ is $\sum_{k=0}^\infty k p_k$.
 Each node has a \emph{screen}, which holds the meme of current interest to that node (see Fig.~1). For simplicity, we assume here that each screen has capacity for only one meme, though this case is easily extended \footnote{See Supplementary Material accompanying this paper.}.
During each time step (with time increment $\Delta t=1/N$), one node is chosen at random.
 With probability $\mu$, the selected node \emph{innovates}, i.e., generates a brand-new meme, that appears on its screen, and is tweeted (broadcast) to all the node's followers.
 Otherwise (with probability $1-\mu$), the selected node (re)tweets the meme currently on its screen (if there is one) to all its followers, and the screen is unchanged. If there is no meme on the node's screen, nothing happens.
 When a meme $m$ is tweeted, the popularity of meme $m$ is incremented by 1 and the memes currently on the followers' screens are overwritten by meme $m$.\\
\begin{figure*}
\centering
\epsfig{figure=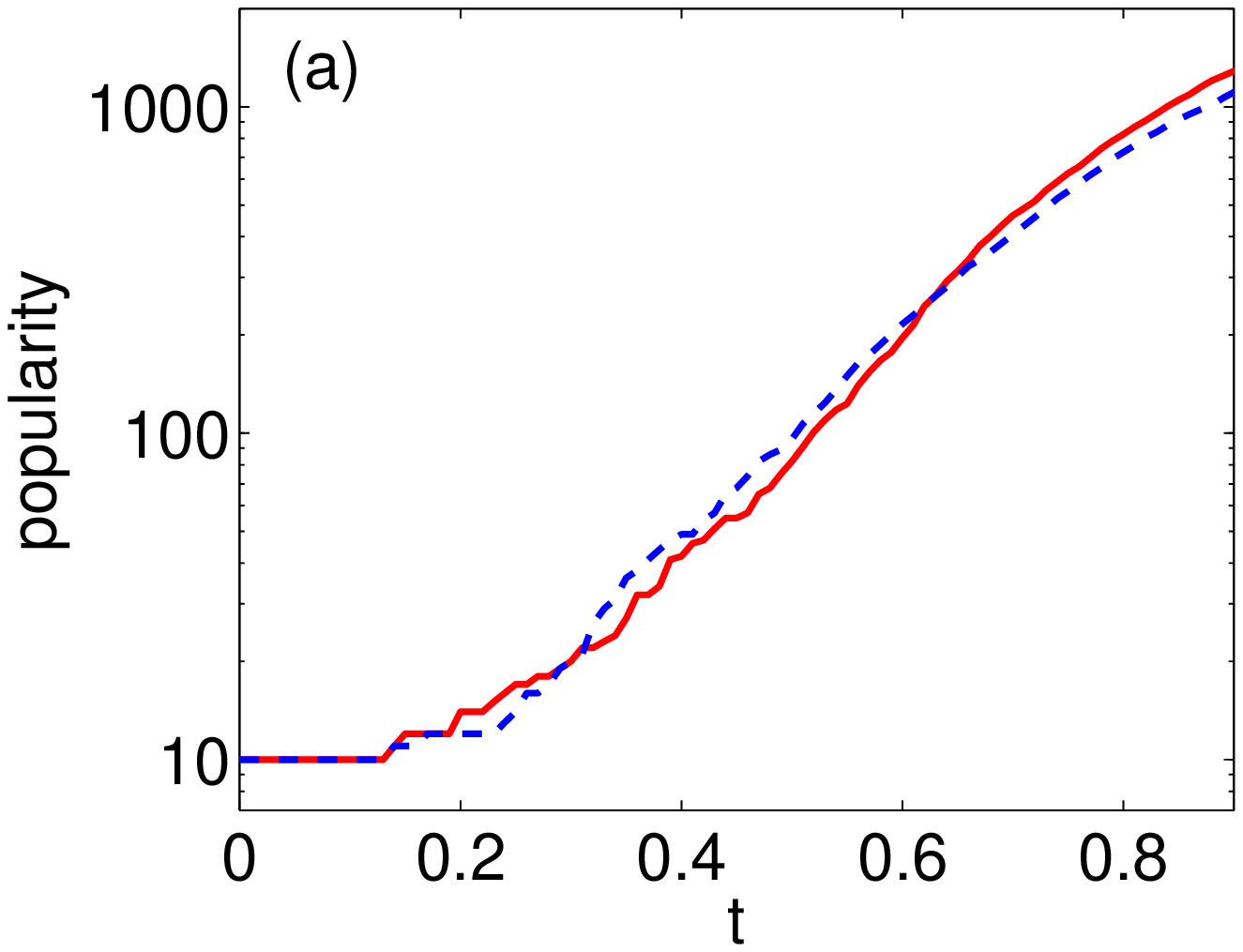,width=5.5cm}
\epsfig{figure=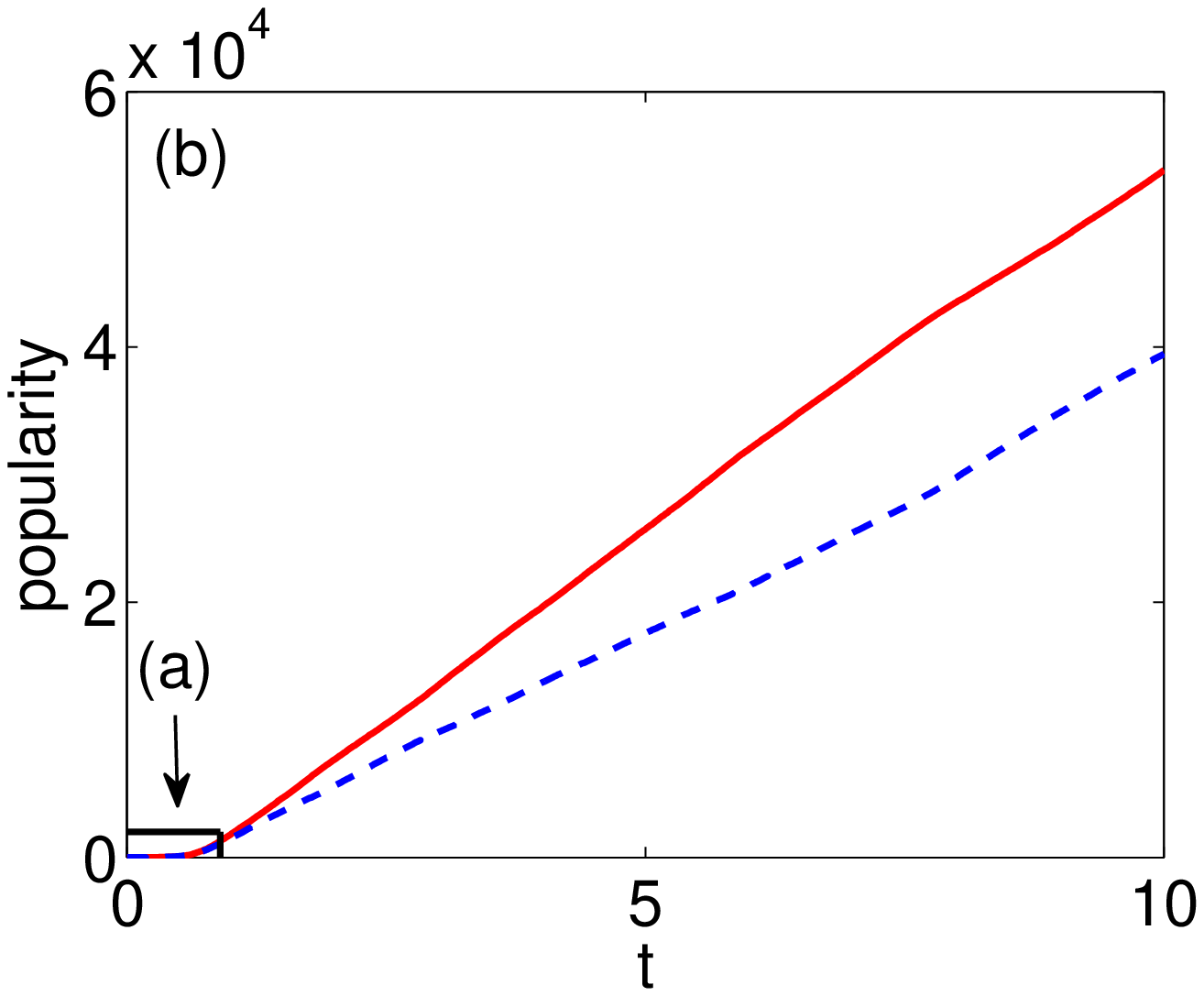,width=5.5cm}
\epsfig{figure=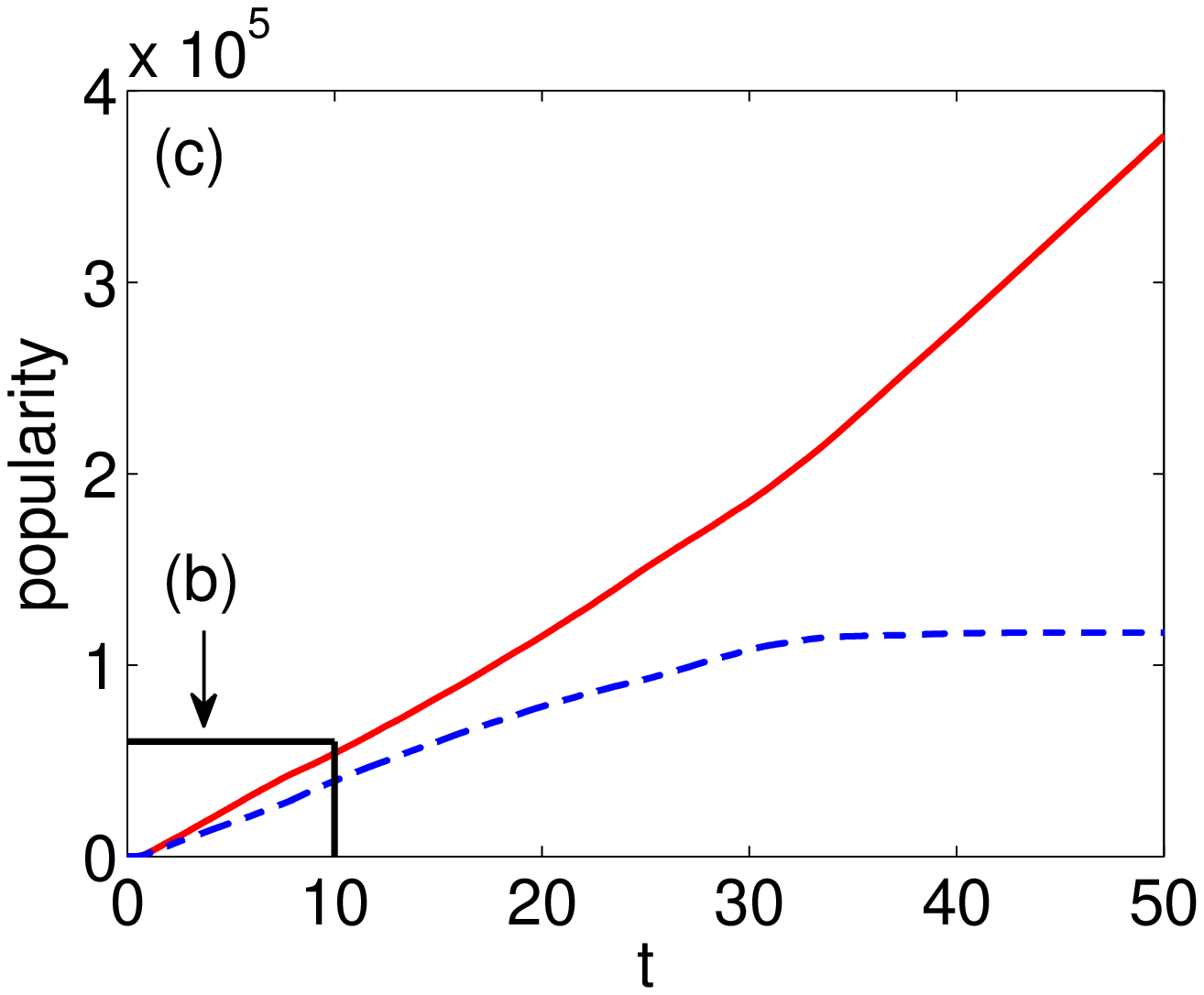,width=5.5cm}
\caption{Popularities $n_r(t)$ and $n_b(t)$ of red and blue memes in a single realization, starting from 10 screens each. Note the different timescales in each figure, and the fact that the vertical scale in (a) is logarithmic. The inset boxes in (b) and (c) show the area of the previous figures.}\label{figfew1}
\end{figure*}

\noindent\emph{Two memes; initially no competition:}
As a first examination of the model's dynamics, we consider just two memes (called red and blue), each of which is initally present on a small number of screens, with every other screen being empty, and with no innovation ($\mu=0$).
%In a typical realization, the red and blue memes each spread quickly onto  empty screens and their popularities both exhibit exponential growth in time (Fig.~\ref{figfew1}(a)). Once most screens are occupied by either the blue or the red meme, the memes compete for space on screens, and both  popularities grow approximately linearly with time (Fig.~\ref{figfew1}(b)). On longer timescales, one of the memes is eventually extinguished from all screens, and its popularity curve saturates (e.g., the blue meme in Fig.~\ref{figfew1}(c)).
A simple mean-field analysis of this two-meme case gives some useful insight. We assume all nodes have $z$ followers, and follow $z$ others, where $z$ is the mean out-degree $\sum_k k p_k$ of the network. Let $r(t)$ be the fraction of screens occupied by the red meme at time $t$, with $b(t)$ the corresponding fraction of blue-meme screens. Since nodes are selected at random to tweet, the expected popularity (i.e., the cumulative number of tweets up to time $t$) for the red meme, $n_r(t)$, is related to $r(t)$ by $d n_r/d t = r(t)$, with a similar relation for the blue meme. Under the mean-field assumptions, a deterministic approximation for $r(t)$ and $b(t)$ is given by the solution of the pair of equations
%\begin{align}
%\frac{d r}{d t }& = - z b r +z r (1-r), \label{1r}\\
%\frac{d b}{d t} & = -z b r + z b(1-b). \label{1b}
%\end{align}
\begin{equation}
\frac{d r}{d t } = - z b r +z r (1-r),\quad
\frac{d b}{d t}  = -z b r + z b(1-b). \label{1r}
\end{equation}
The first term on the right-hand-side of the first equation, for example, accounts for a decrease in the number of red-meme screens due to memes being overwritten by blue-tweeting nodes. This occurs when a blue meme is tweeted (with probability $b(t)$ in a given time step), and affects a fraction $r(t)$ of the $z$ followers of the tweeting node, giving the term $-z b r$. The second term describes the growth of red memes due to a red meme tweeting (with probability $r(t)$) to non-red followers, the expected number of which is $z(1-r(t))$.

Equations (\ref{1r})  can be solved analytically: the fraction of non-empty screens is $i(t)=r(t)+b(t)$, with $i(0)\ll 1$, and its dynamics obey the logistic differential equation $d i/d t=z i (1-i)$, which is precisely the mean-field approximation for the infected population fraction in a susceptible-infected (SI) epidemic model. When   $r(t)$ and $b(t)$ are both very small the solutions show exponential growth in screen occupation, and hence in the accumulated tweets (i.e., popularities) $n_r(t)$ and $n_b(t)$---see Fig.~\ref{figfew1}(a)---similar to early-stage growth of independent diseases \cite{Barratbook}. The exponential growth continues until $i(t)$ is of order 1, by which time most screens show either the red or the blue meme. When $r(t)+b(t)=1$, the right hand sides of Eqs.~(\ref{1r})  are both zero. This means that---under the mean-field assumptions that give this deterministic limit---the numbers of screens showing each meme remain constant thereafter, and so the popularities $n_r(t)$ and $n_b(t)$ grows linearly in time, as in Fig.~\ref{figfew1}(b). This balance is a dynamic one, as the two memes continue to compete for the resource of screen space, but the rate of growth for each meme is precisely equal to the rate of loss due to being overwritten by the other meme. Thus the linear growth in popularity is induced by the competition between memes, in contrast to the exponential growth at earlier times (Fig.~\ref{figfew1}(a)) when the memes were not competing for the same resources \cite{Forgerini11,Szabo10}.

The mean-field approximation used above ignores finite-$N$ effects, which cause stochastic fluctuations in the number of screens about the mean values $r(t)$ and $b(t)$. In the long-time limit, it is these fluctuations that eventually lead to one meme becoming extinct, with the other filling all screens (as in Fig.~\ref{figfew1}(c)). Stochastic fluctuations are also important at early times, when there are only very few screens showing either meme. In order to model the important role of stochastic fluctuations, and also to examine how the results presented here extend to cases with very many memes, we next consider  a heavily-competitive environment containing multiple memes.\\
%, using branching process theory \cite{Harrisbook}.

\noindent\emph{Multiple competing memes:}
{ Now suppose that there are no empty screens in the network---so we are in the highly-competitive regime}---and the innovation probability $\mu$ may be non-zero. Competition between memes for the limited resource of user attention (i.e., screen space) leads naturally to some memes becoming extremely popular, while others are only moderately popular, or are ignored.
 %We are interested in the distribution of popularity across the set of all $N$ memes \emph{initially} present in the system.
%, motivated by, for example, the cascade size distributions found empirically for the spread of URLs %through Twitter (see Fig.~4(a) of Bakshy et al, Fig.~6(d) of Lerman et al).
We show that the model produces fat-tailed distributions of popularity, which are power-law in the limit $\mu\to 0$. This is explained using a branching process description of the model, where the competitive environment causes each meme to follow a critical branching process (for which power-law distributions are expected \cite{Adami02,Goh03}).

%\subsection{Branching process approximation}
%\begin{figure*}
%\centering
%\epsfig{figure=fig2a.eps,width=5.5cm} \hspace{1cm}
%\epsfig{figure=fig2b.eps,width=5.5cm}
%\caption{Complementary cumulative distribution functions (CCDFs)---the fraction of memes with popularity $\ge n$---for numerical simulations (with $N=10^5$), compared with the theory of Eq.~(\ref{2}). In this network every node has exactly $z=10$ followers, $p_k=\delta_{k,10}$. In (a), the innovation parameter $\mu$ is zero; in (b), $\mu=0.02$, causing a large-size cutoff as predicted by Eq.~(\ref{4}). In (b), only the popularities of the memes present at initialization are shown. Dashed lines correspond to CCDFs for power law popularity distributions $P_n\propto n^{-\alpha}$.}\label{figz10}
%\end{figure*}
The branching process description is strictly valid only when the number of screens occupied by { a} single meme is a small fraction of $N$, but we note that this is the case for long epochs of {time in} a competitive environment with many memes. We assume here that all nodes follow (approximately) $z$ other nodes, so the in-degree distribution is homogeneous, but we consider heterogeneous  distributions of out-degrees. Before examining the details of the branching process, it is worth highlighting the source of criticality in the model when $\mu=0$. In a single time step $\Delta t$, a tweeting node creates (or ``gives birth to'') an average of $z \Delta t$ new copies of the meme on its screen by overwriting the screens of its followers. However, each screen can be overwritten by another meme (causing ``death'' of the overwritten meme) with probability $z \Delta t$, and so the birth and death rates of memes are, on average, exactly balanced, giving a critical branching process. This balance between births and deaths remains critical when the model is enhanced in several ways{, including modifying the rules so that nodes retweet any given meme at most once, see Sec.~S5 of [47]}.

Next we give details of the branching process description of the model. We denote the distribution of popularities at age $a$  by $q_n(a)$: this is the probability that a meme has been tweeted $n$ times when its age is $a$ (i.e., at a time $t_b+a$, where $t_b$ is the birth time of the meme). This distribution  can be represented via its probability generating function (PGF) \cite{Wilfbook,Newman01} $H(a,x)$, defined by $H(a,x)\equiv \sum_{n=1}^\infty q_n(a) x^n$. The network topology is described by the PGF for the out-degree distribution: $f(x)\equiv \sum_{k=0}^\infty p_k x^k$. The mean degree  is $z=f'(1)$ and we assume all nodes have in-degree $z$.

\begin{figure*}
\centering
\epsfig{figure=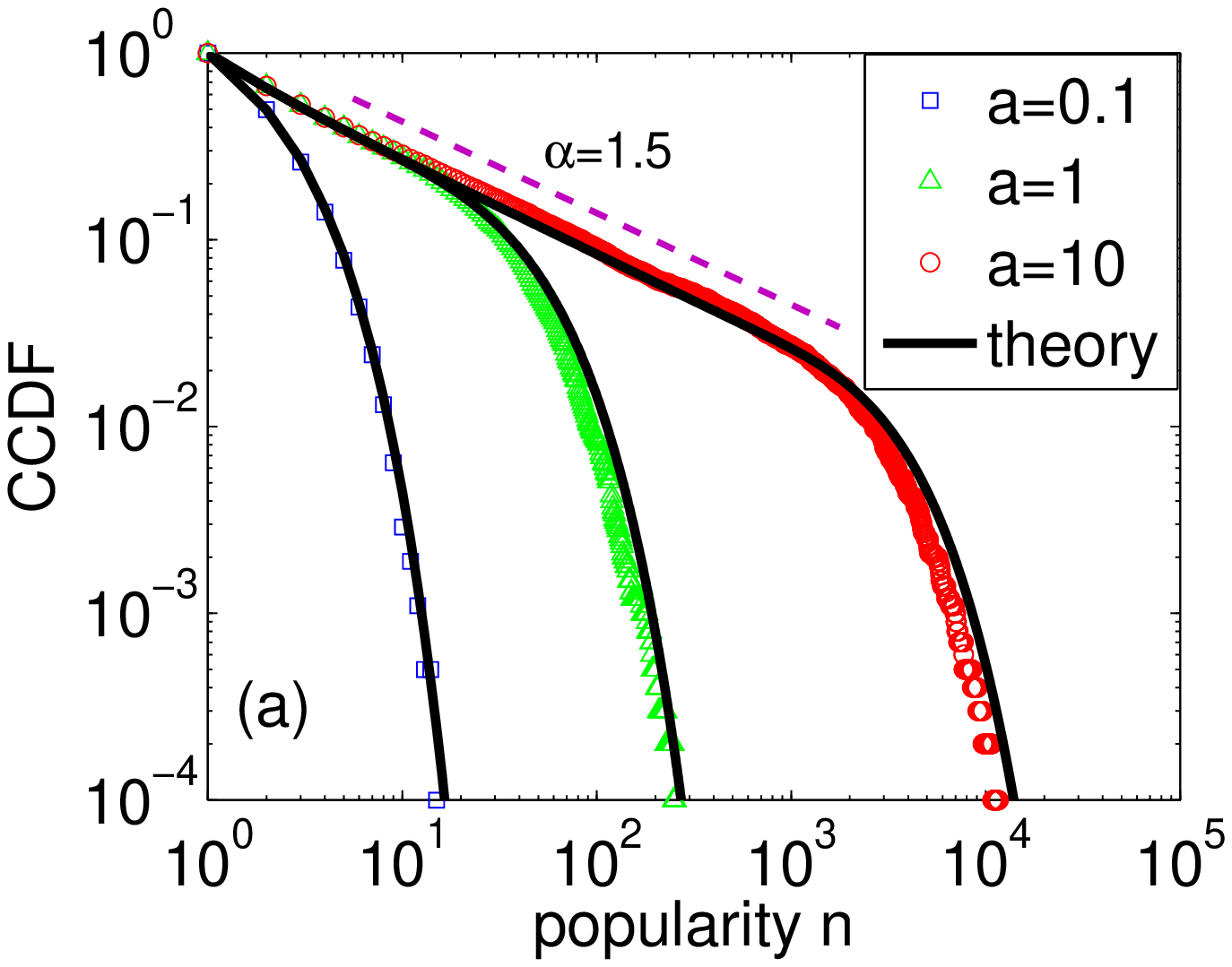,width=5.5cm}
\epsfig{figure=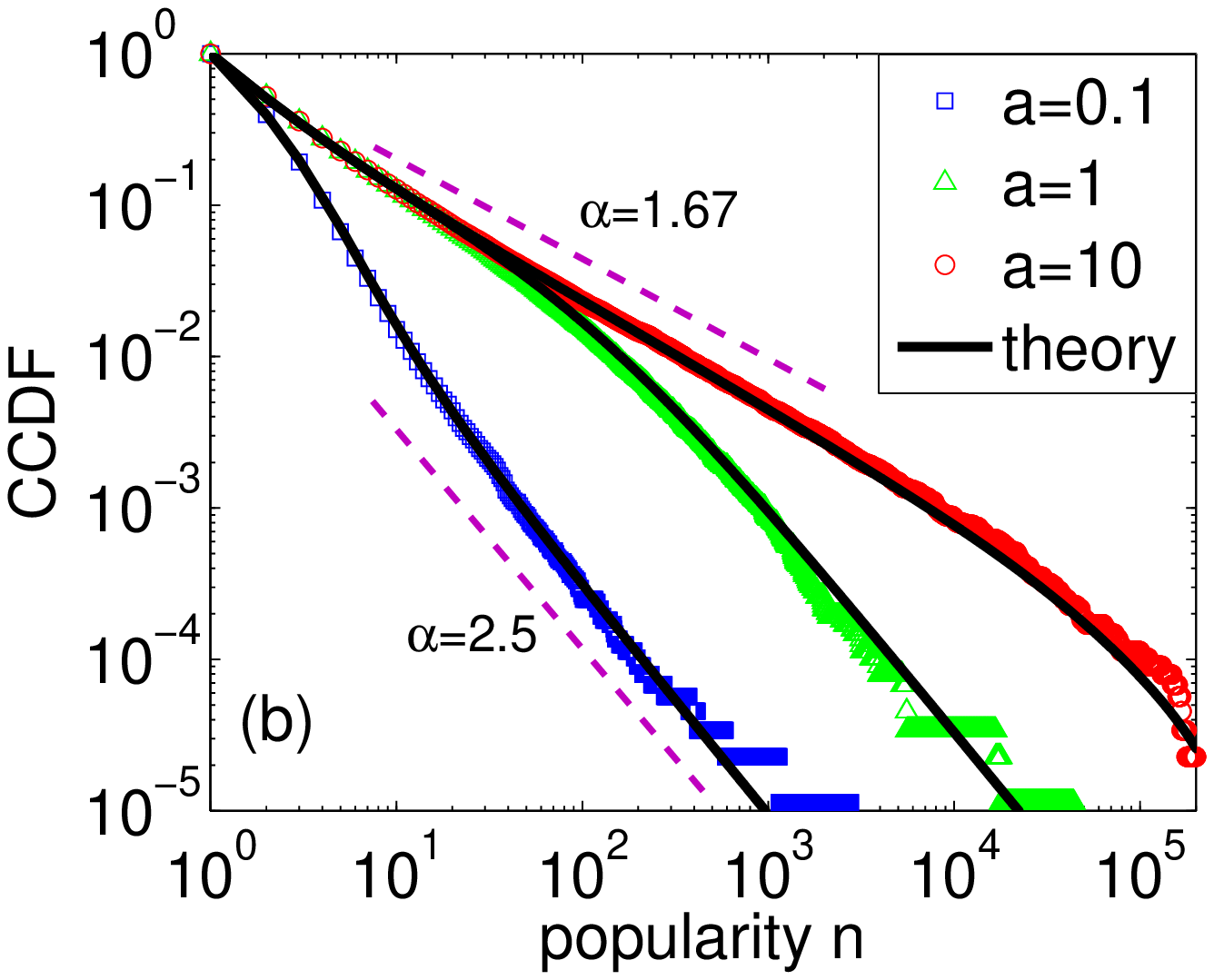,width=5.5cm}
\epsfig{figure=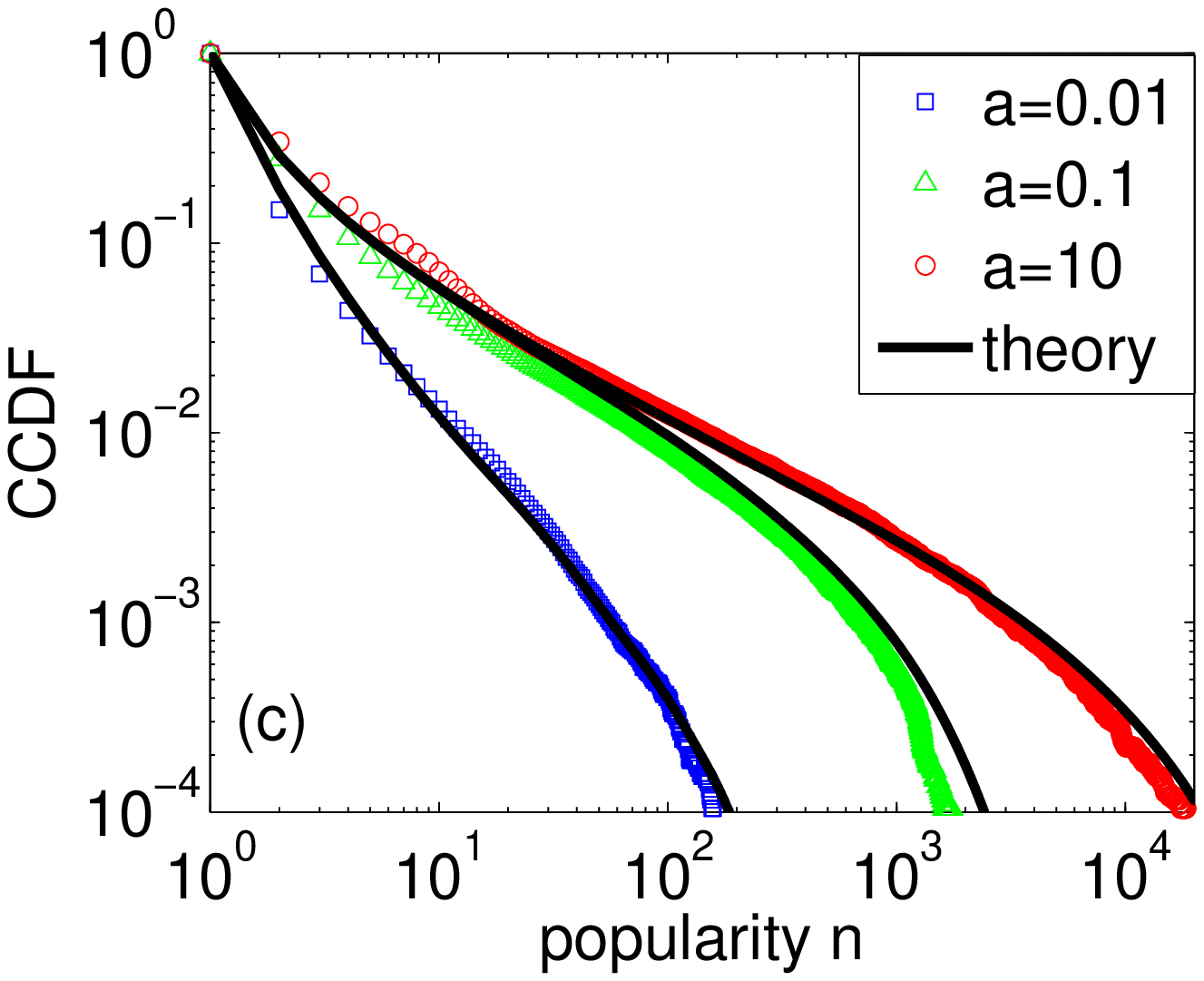,width=5.5cm}
\caption{{
Complementary cumulative distribution functions (CCDFs)---the fraction of memes with popularity $\ge n$---for numerical simulations, compared with the theory of Eq.~(\ref{2}). Dashed lines correspond to CCDFs for power law popularity distributions $P_n\propto n^{-\alpha}$.
(a) $p_k=\delta_{k,10}$, $N=10^5$, $\mu=0$. (b) $p_k\propto k^{-\gamma}$ for $k\ge 4$ with $\gamma=2.5$  (mean degree $z=10.6$), $N=10^6$, $\mu=0.01$.  (c) Twitter network of the Spanish 15M movement \cite{Borge11,Gonzalez11},   $N=87,569$, $z=69$, $\mu=0.05$.
  }}\label{figSFN}
\end{figure*}

To calculate $q_n(a)$, we first find $H(a,x)$ and then employ an inversion technique based on Fast Fourier Transforms (FFTs) \cite{Cavers78,Marder07,Noel12}. It proves convenient to introduce $G(a,x)$, defined as the PGF for the \emph{excess popularity} distribution at age $a$  of memes  that originate from a single randomly-chosen screen (the \emph{root} of the tree).
{ The tweet event that creates the root is not counted by $G$: this event increases the popularity of the meme by 1, and places the meme upon the root screen and the screens of all followers of the root node. Consequently, the PGF for the popularity of age-$a$ memes is given by
$H(a,x) = x \, G(a,x)\, f(G(a,x))$.}
In Sec.~S1 of [47] we derive the following ordinary differential equation for $G(a,x)$, parameterized by $x$:
\begin{equation}
\frac{\partial G}{\partial a} = z+\mu-(z+1)G+(1-\mu) x\, G\, f(G) \label{2}.
\end{equation}
This equation is easily solved using standard numerical methods, starting from the initial condition $G(0,x)=1$. Some analysis is also possible [47]: the mean popularity $\partial H/\partial x (a,1)$, for example, grows linearly with age until $a$ is of the order $1/\mu(z+1)$, thereafter it saturates at a value of $1/\mu$. By expanding $G(a,x)$ as a Taylor series about $x=0$, the probabilities $q_n(a)$ for low $n$ may be determined explicitly.
%For example, the probability that a meme's popularity never grows beyond its initial value of $n=1$ is given by the large-$a$ limit of $q_1$, which is $(z+\mu)/(z+1)$, showing that (for $\mu \ll 1$ and $z \gg 1$) most memes do not go viral \cite{Goel12, Banos13} as they are never retweeted.
The popularity distribution for larger values of $n$ are determined in a computationally efficient manner using FFTs \cite{Cavers78,Marder07,Noel12}: our implementation (Sec.~S2 of [47]) determines probabilities $q_n$  for $n$ values up to several thousand,  shown as black curves in
 Fig.~\ref{figSFN} \footnote{Octave/Matlab codes for solving Eq.~(\ref{2}) and inverting the generating functions
 are available
 for download from \texttt{www.ul.ie/sdcs/people/kevin-osullivan}.}.
% from the authors upon request.}.
The coloured symbols are the results of stochastic simulations of the model,
  %on networks of size $N=10^5$; results for two independent realizations of each network are shown, each
  giving  popularity distributions for memes at various ages $a$. The match between theory and simulation is very good.
 Figure~\ref{figSFN}(a) shows the popularity distributions  on networks where each node has exactly $z=10$ followers, while
   %in the cases $\mu=0$ (Fig.~\ref{figz10}(a)) and $\mu=0.02$ (Fig.~\ref{figz10}(b)).
   Fig.~\ref{figSFN}(b) is for a network where the number of followers (out-degree of a node) has a power-law distribution: $p_k \propto k^{-\gamma}$ for $k\ge 4$, with $\gamma=2.5$ (and $p_k=0$ for $k<4$). In both cases the $k$ followers of a given node are assigned at random, so the in-degree distributions are Poisson.
  %The corresponding generating function $f(x)$ for this case is $\text{Li}_\gamma(x)/\zeta(\gamma)$ where $\text{Li}_\gamma$ is the polylogarithm function of order $\gamma$ and $\zeta$ is the Riemann zeta function \cite{Goh03}.
  % We observe power-law popularity distributions  $q_n\propto n^{-\alpha}$, with various exponents $\alpha$, and with an exponential cut-off in Fig.~\ref{figz10}.

%[JG: Why does asymptotic slope look different from theory curve at largest $n$?]
% cf inv_gen_fns_8k.nb

The observed power-law popularity distributions can be understood using
the long-time (or ``old-age'', $a\to \infty$) asymptotics of the branching process, found by analyzing the limiting solutions of Eq.~(\ref{2}) in the complex $x$-plane (Sec.~S3 of [47]);
  we summarize the main results as follows.
If the out-degree distribution $p_k$ has a finite second moment (i.e., if $f^{\prime\prime}(1)<\infty$), then the $a\to \infty$ limit of the popularity distribution has the asymptotic form
    $
    %\begin{equation}
    q_n(\infty) \sim
     A \,n^{-\frac{3}{2}} e^{-\frac{n}{\kappa}}%\quad
     \text{ as }n\to \infty,%\label{4}
    %\end{equation}
    $
    where
    $
    \kappa=\frac{2(f^{\prime\prime}(1)+2 z)}{\mu^2(z+1)^2}
    $
    and $A=(z+1)\left[ 2 \pi (f^{\prime\prime}(1)+2 z)\right]^{-\frac{1}{2}}$.
    This formula shows that the popularity distribution is of power-law form $n^{-{3/2}}$, up to an exponential cutoff at  $n\approx \kappa$. However, the cutoff size $\kappa$ limits to infinity as the innovation rate $\mu$ goes to zero, and $\kappa$ can be large even for non-zero $\mu$ if the second moment of the distribution $p_k$ is large (since $f^{\prime\prime}(1)=\sum_k k(k-1)p_k$).

 If $p_k \propto k^{-\gamma}$ for large $k$ with $2<\gamma<3$, then $f^{\prime\prime}(1)$ is infinite, and a different asymptotic analysis is required. In this case we find, similar to \cite{Goh03}, that as $n\to\infty$,
    \begin{equation}
    q_n(\infty)\sim \left\{ \begin{array}{cl}
     B \, n^{-\frac{\gamma}{\gamma-1}} & \quad \text{ if $\mu=0$},\\
     C\, n^{-\gamma} & \quad \text{   if $\mu>0$,}
     \end{array}\right. \label{eqn4}
    \end{equation}
   with prefactors $B$ and $C$ given in [47].
   % Kevin's fixes implemented 3Apr13
  %$
  %  B(\gamma)  = \frac{1}{\pi}\sin\left(\frac{\pi}{\gamma-1}\right) \Gamma\left(\frac{\gamma}{\gamma-1}\right)\left[\frac{\zeta(\gamma)}{\Gamma(1-\gamma)}\right]^\frac{1}{\gamma-1}
  %$
  %and
  %$
  %  C(\gamma)  = %    \frac{1}{\pi}\sin\left(\pi\gamma\right)\frac{\Gamma(1-\gamma)\Gamma(\gamma)}{\zeta(\gamma)}\left[\frac{1-\mu}{\mu(z+1)}\right]^\gamma,
%\frac{1}{\zeta(\gamma)}\left[\frac{1-\mu}{\mu\left(1+\zeta(\gamma-1)/\zeta(\gamma)\right)}\right]^\gamma,
%$
%where $\Gamma$ is the gamma function.
%, and $z=\zeta(\gamma-1)/\zeta(\gamma)$.
Thus in the zero-innovation limit, the popularity distribution has a power-law exponent $\gamma/(\gamma-1)$ that is smaller than the exponent $\gamma$ of the out-degree distribution.
 %If $\mu>0$, the large-$n$ power-law exponent for the popularity distribution is the same as that of the out-degree distribution, but numerical results (Fig.~\ref{figSFN}(b)) show that for moderately large $n$ the distribution still decreases more slowly that the out-degree distribution.

{Figure~\ref{figSFN}(c) compares theory and simulation results for the model  on the real Twitter network of \cite{Borge11,Gonzalez11}. The theory matches the simulation results rather well, despite the fact that this network is not tree-like---indeed, 44\% of links are reciprocal links---
and does not have a homogeneous in-degree distribution, as assumed in the derivation of the theory. The accuracy of results from tree-based theories applied to real-world networks has been noted previously \cite{Gleeson12} and is examined further for this model in Sec.~S4 of [47].}\\

\noindent\emph{Conclusions:} We have used a simple model of meme diffusion to illustrate the phenomenon of competition-induced criticality. It is straightforward to generalize the basic model and the derivation of Eq.~(\ref{2})---for example, by: (i) increasing the capacity of screens to $c>1$ memes,  (ii) allowing followers to reject a meme tweeted to them with some probability so it does not appear on their screen, { or (iii) permitting nodes to retweet a meme at most once}---and to show that the CIC property is retained in the more general cases [47].
Despite their simplicity, we believe that the understanding of such analytically tractable models provides important insights on the origin of regularities observed in empirical data. For instance, our model does not include fat-tailed distributions of in-degrees, user activity levels, or response times \cite{Weng12,Lerman12,Iribarren11}---these will be added in future work---but it can nevertheless produce fat-tailed popularity distributions.
%These insights will guide the eventual creation of parsimonious but accurate models of human choice dynamics that can reproduce key characteristics of the rapidly expanding range of empirical data from online social networks.

{We have also shown that the CIC model produces avalanches of popularity whose sizes  have the same steady-state distributions as those found in sandpile models of self-organized criticality \cite{Goh03}. This is intriguing because the CIC mechanism is quite distinct from the sandpile paradigm: nodes do not have thresholds for triggering avalanches, for example, and the CIC popularity avalanches evolve on the same timescale as the general dynamics. We speculate that competition for limited resources may therefore play an important role in many other application areas where SOC-like phenomena have been identified in experiments or numerical simulations \cite{Krider97,Arenas02,Sprott04,Newman96,Sole96,Haimovici13,Rubinov11}.}\\

\noindent\emph{Acknowledgements:}
This work was partially funded by Science Foundation Ireland (11/PI/1026 and 09/SRC/E1780), the Engineering and Physical Sciences Research Council (MOLTEN, EP/I016058/1)  and
by the FET-Proactive project PLEXMATH. We thank Peter Fennell for assistance with Fig.~3(c) and acknowledge helpful discussions with D.~Cellai, S.~Melnik, M.~A.~Porter, J-P Onnela and F.~Reed-Tsochas.
We acknowledge the SFI/HEA Irish Centre for High-End Computing (ICHEC) for the provision of computational facilities, and the COSNET Lab for publishing the 15M dataset.

\clearpage
\section*{SUPPLEMENTARY MATERIAL}
\renewcommand{\thefigure}{S\arabic{figure}}
\setcounter{figure}{0}
\renewcommand{\thetable}{S\arabic{table}}
\setcounter{table}{0}
\renewcommand{\theequation}{S\arabic{equation}}
\setcounter{equation}{0}

\section{S1 Derivation of Equation (2)}

In the main text we introduced $G(a,x)$ as the probability generating function (PGF) of the excess popularity distribution, and $H(a,x)$ as the PGF of the popularity distribution, with
\begin{eqnarray}
H(a,x) &=& x\, G(a,x)\, f(G(a,x)) \nonumber\\
&=& \sum_{n=0}^\infty q_{n}(a)x^n,
\end{eqnarray}
where $q_n(a)$ is the probability that a meme of age $a$ has  popularity $n$. It proves convenient here to also introduce $G^{(k)}(a,x)$, the PGF of the excess popularity distribution for a meme seeded by (i.e., first tweeted by) a node with out-degree $k$; we call this node's screen the \emph{root screen} of the retweet-cascade tree. Using the out-degree distribution $p_k$ of the network, we have the relation
\begin{equation}
G(a,x) = \sum_k p_k G^{(k)}(a,x).
\end{equation}
\begin{figure}
%\centering
\begin{minipage}{3in}
\hspace{-13.9cm}
\epsfig{figure=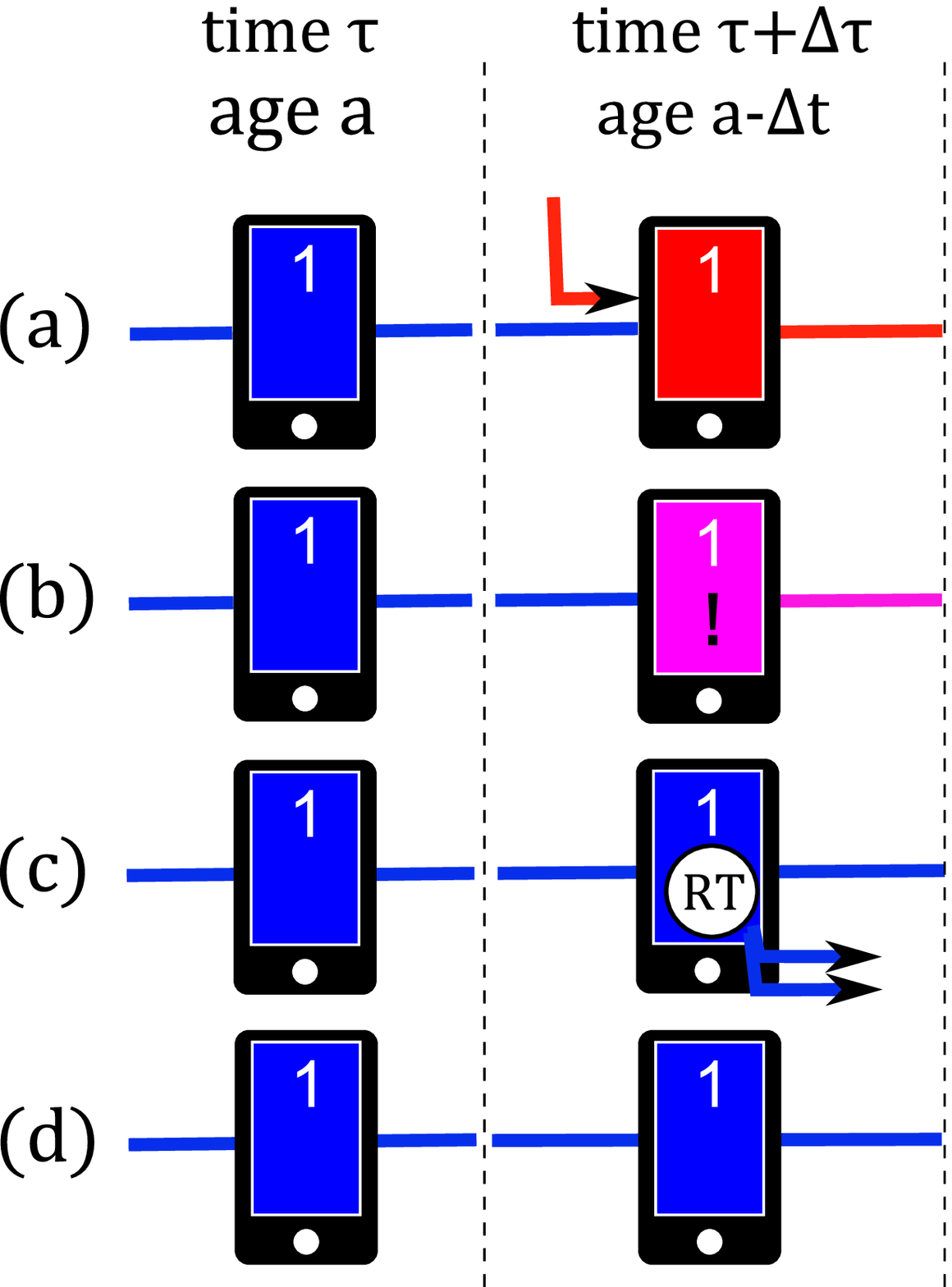,width=4.6cm}
%\caption{caption}\label{figS1}
 \end{minipage}
%\begin{table}
%\begin{center}
\hspace{-8.2cm} %\quad % \qquad
\begin{minipage}{2in}
\vspace{-2.0cm}
\begin{tabular}{ccc}
%\hline
\vspace{-0.4cm}\\{\bf Outcome for screen $\mathbf{S_1}$ }& \bf{ Probability }&\bf{ Contribution to $\mathbf{G^{(k)}(a,x)}$}\\
\vspace{0.55cm}\\
%\hline
\vspace{0.9cm}
Overwritten & $z\, \Delta t$ & 1\\
\vspace{0.9cm}
Selected, innovates & $\mu\, \Delta t$ & 1\\
\vspace{0.9cm}
Selected, retweets & $(1-\mu)\, \Delta t$ & $x G^{(k)}(a-\Delta t,x) [G(a-\Delta t,x)]^k$\\
\vspace{-1.6cm}
Not selected, survives & $1-(z+1) \Delta t$ & $G^{(k)}(a-\Delta t,x)$\\
\end{tabular}
%\end{center}
%\caption{Parameter values for the fitting functions $m(a)$ used in Fig.~\ref{figSI2}.}
%\label{tabSI1}
%\end{table}
\end{minipage}
\caption{Summary of the single-timestep outcomes that contribute to the PGF $G^{(k)}(a,x)$.}\label{figS2}
\end{figure}
The meme on the root screen is the root of the cascade tree that results from the meme being tweeted and  subsequently retweeted over a period of time, see Fig.~1 of the main text. If we fix a time $t=\Omega$ as the observation time for the cascade sizes, then $G(a,x)$ is the PGF for the sizes (as observed at time $\Omega$) of trees that are rooted at time $\Omega-a\equiv \tau$. We derive a relation between the PGF for the sizes of trees at age $a$ (i.e., those rooted at time $\Omega-a$) and the PGF for tree sizes at age $a-\Delta t$ (i.e., those rooted at time $\Omega-a+\Delta t$), as follows.

 Consider a meme on a given screen (call this screen $S_1$), at time $t=\tau$ : this is the root of the tree we call $\text{Tree}(S_1)$, which has age $a$ at the observation time $t=\Omega$; let $k$ be the out-degree of the node with screen $S_1$. At the next time step $t=\tau + \Delta t$, there are four possible outcomes for this particular screen that contribute to the PGF $G^{(k)}(a,x)$, refer to Fig.~\ref{figS2}:
 \begin{itemize}
 \item Outcome (a): the screen $S_1$ is overwritten by some other meme that is tweeted by another node. This terminates $\text{Tree}(S_1)$---setting the corresponding generating function  to 1---as no future tweets can now result from the chosen root. Outcome (a) occurs with probability $z \Delta t$, since a node follows, on average, $z$ other nodes, each of which is the tweeter with probability $1/N=\Delta t$. So outcome (a) contributes $z \Delta t$ (the probability of occurrence multiplied by the resulting generating function term) to the PGF $G^{(k)}(a,x)$. (Note that all terms of order $(\Delta t)^2$ and higher are ignored here and below, as these are negligible when we take the limit $\Delta t \to 0$.)
 \item  Outcome (b): the screen $S_1$ is selected as the updater in the current time step (with probability $\Delta t$) and innovates (with probability $\mu$), so terminating $\text{Tree}(S_1)$. The contribution to $G^{(k)}(a,x)$ is thus $\mu \Delta t$.
 \item Outcome (c): the screen $S_1$ is selected for update (probability $\Delta t$) and retweets its meme (probability $1-\mu$). This (i) adds one to the size of $\text{Tree}(S_1)$, whilst (ii) the branch on screen $S_1$ survives another time step, becoming the origin of a new tree rooted at $\tau+\Delta t$, which has age $a-\Delta t$ at the observation time. Moreover, (iii) screen $S_1$ is the parent of $k$ new branches of $\text{Tree}(S_1)$: each new branch acts as the root of a tree with PGF $G(a-\Delta t,x)$ (since the out-degrees of the newly-spawned roots are random). These effects (i)--(iii) lead to generating function contributions of $x$, $G^{(k)}(a-\Delta t,x)$, and $[G(a-\Delta t,x)]^k$, respectively, and since these occur simultaneously, the overall contribution of outcome (c) to $G^{(k)}(a,x)$ is $(1-\mu)\Delta t\, x \, G^{(k)}(a-\Delta t,x) [G(a-\Delta t,x)]^k$.
\item Outcome (d): the survival of $\text{Tree}(S_1)$, with none of the other  outcomes (a)--(c) occurring: the probability of this is $1-(z \Delta t + \mu \Delta t +(1-\mu)\Delta t) = 1-(z+1)\Delta t$, and the screen can then be considered as the origin of a new tree that is rooted at time $\tau+\Delta$, and so has age $a-\Delta$ at the observation time $\Omega$. The overall contribution to $G^{(k)}(a,x)$ from this outcome is therefore $(1-(z+1)\Delta t) G^{(k)}(a-\Delta t,x)$.
\end{itemize}
Putting all four outcomes together gives an expression for $G^{(k)}(a,x)$, correct to first order in $\Delta t$:
\begin{align}
G^{(k)}(a,x) &=\nonumber \\
&\hspace{-1cm}\underbrace{z\, \Delta t}_{\text{(a)}}+\underbrace{\mu\, \Delta t}_{\text{(b)}}+\underbrace{(1-\mu)\Delta t\, x\, G^{(k)}(a-\Delta t,x)\, [G(a-\Delta t,x)]^k}_\text{(c)} +\underbrace{(1-(z+1)\Delta t)G^{(k)}(a-\Delta t,x)}_{\text{(d)}}\label{Gkeqn0},
\end{align}
and taking the limit $\Delta t \to 0$ yields an ordinary differential equation for $G^{(k)}(a,x)$, parameterized by $x$:
\begin{equation}
\frac{\partial G^{(k)}}{\partial a} = z+\mu-(z+1)G^{(k)}+(1-\mu) x\, G^{(k)}\, [G]^k \label{Gkeqn}.
\end{equation}
Averaging over the possible out-degrees of the root node---by multiplying by $p_k$ and summing over all $k$---gives the following equation for $G(a,x)$:
\begin{equation}
\frac{\partial G}{\partial a} = z+\mu-(z+1)G+(1-\mu) x\, \sum_k p_k \,G^{(k)}\, [G]^k \label{Gkeqn2}.
\end{equation}
Solving this equation for $G$ requires also finding the functions $G^{(k)}$, for all $k$. However, if we make the following approximation
\begin{equation}
\sum_k p_k \left(G^{(k)}\, [G]^k\right)  \approx \left(\sum_k p_k \,G^{(k)}\right)\left(\sum_k p_k\, [G]^k \right) = G f(G), \label{momclosure}
\end{equation}
we obtain the single differential equation for $G(a,x)$ given by Eq.~(2) of the main text. The simplifying moment-closure assumption (\ref{momclosure}) will be examined in detail in further work; our numerical simulations  indicate that it leads to quite accurate results for networks with reasonably large mean degree $z$ (e.g. $z\approx 10$ as in Fig.~3(a) and (b) of the main text).

It is straightforward to generalize the derivation above to allow each node's screen to have capacity $c\ge 1$, meaning that the screen can simultaneously hold $c$ memes: we then consider each screen to be composed of $c$ distinct \emph{screen-slots}. When retweeting, a node (user) chooses one of their $c$ screen-slots at random to determine the meme that is transmitted to its followers; if the chosen screen-slot is empty then nothing happens. When a node innovates, or when it receives a meme from another node, the new meme is placed in a randomly-chosen screen-slot, overwriting  any existing meme in that slot. An additional generalization is to allow for tweeted memes to be accepted onto followers' screens with probability $\lambda\le 1$ (with $\lambda=1$ giving the base case of the main text).  Incorporating these generalizations into the derivation above leads to the following equation for $G(a,x)$:
\begin{equation}
c \frac{\partial G}{\partial a} =\lambda z+\mu-(\lambda z+1)G+(1-\mu) x\, G\, f(1-\lambda+\lambda G) \label{Gen},
\end{equation}
which reduces to Eq.~(2) of the main text in the case $c=1$ and $\lambda=1$; the corresponding equation for $H(a,x)$ is $H=x \,G\, f(1-\lambda+\lambda G)$.
 We use this more general case throughout the Supplementary Material to demonstrate that the competition-induced criticality phenomenon is robust to changes in the model.

The mean popularity of age-$a$ memes,
\begin{equation}
m(a) \equiv \sum_{n=1}^\infty n q_n(a) =  \frac{\partial H}{\partial x}(a,1) = 1+(\lambda z+1) \frac{\partial G}{\partial x}(a,1),
\end{equation}
can be found found by differentiating Eq.~(\ref{Gen}) with respect $x$ to obtain the linear equation
\begin{equation}
c \frac{d m}{d a} = (\lambda z+1)(1- \mu m),
\end{equation}
with $m(0)=1$. The solution is
\begin{equation}
m(a) = \left\{ \begin{array}{cl}
1+ \frac{\lambda z +1}{c} a & \text{ if } \mu=0\\
\frac{1}{\mu}- \frac{1-\mu}{\mu}e^{-\frac{\mu(\lambda z +1)}{c}a} & \text{ if } \mu>0.
\end{array} \right. \label{meana}
\end{equation}
Higher-order moments of the popularity distribution $q_n$ can be obtained similarly by repeated differentiation of Eq.~(\ref{Gen}). It is also possible to directly determine the probabilities $q_n(a)$ for small values of $n$, by expanding $G(a,x)$ as a Taylor series about $x=0$. Setting $x=0$ in Eq.~(\ref{Gen}), for example, immediately yields a closed equation for $G(a,0)$:
\begin{equation}
c \frac{d G(a,0)}{d a} = \lambda z + \mu - (\lambda z+1) G(a,0),
\end{equation}
with solution
\begin{equation}
G(a,0) = \frac{\lambda z+\mu + (1-\mu) e^{-\frac{\lambda z+1}{c} a}}{\lambda z +1}.
\end{equation}
The value of $q_1(a)$---the probability that a meme of age $a$ remains at popularity $n=1$ (i.e., has not been retweeted a time $a$ after its initial tweeting) is then
$
q_1(a)=G(a,0) f(1-\lambda+\lambda G(a,0))
$
and  as $a\to \infty$, $q_1(a)$ approaches the value
\begin{equation}
\frac{\lambda z + \mu}{\lambda z +1 } f\left(1-\lambda +\lambda \frac{\lambda z + \mu}{\lambda z +1 }\right).
\end{equation}
This is the  fraction  of memes that  are forgotten before they are retweeted even once and so their popularity remains at its initial value of $n=1$ forever \cite{Goel12,Banos13}.

\section{S2 Inverting PGFs using Fast Fourier Transforms}

The probability $q_n(a)$ that a meme has popularity $n$ at age $a$ may be determined from the PGF  $H(a,x)$ by repeated differentiation:
\begin{equation}
q_n(a) = \frac{1}{n!}\left.\frac{d^n}{d x^n} H(a,x)\right|_{x=0},
\end{equation}
where $H$ is obtained as $x\, G(a,x)\, f(G(a,x)$ from the (numerical) solution of Eq.~(\ref{Gen}).
However, numerical differentiation is inaccurate for large values of $n$, so we invert the PGF using contour integration in the complex $x$-plane \cite{Cavers78,Newman01}. The inversion integral is given by Cauchy's theorem
\begin{equation}
q_n(a) = \frac{1}{2 \pi i} \oint_C H(a,x) x^{-(n+1)} \, dx,
\end{equation}
where all poles of $H(a,x)$ must lie outside the contour $C$; a common choice for $C$ is the unit circle \cite{Newman01}. Writing $x=e^{i \theta}$ gives the form
\begin{equation}
q_n(a) = \frac{1}{2}\int_{-\pi}^\pi H(a,e^{i\theta})e^{-i n \theta} d\theta,
\end{equation}
 and numerical integration using the trapezoidal rule with $M$ points yields the approximate formula
\begin{equation}
q_n(a) \approx \frac{1}{M} \sum_{m=0}^{M-1} H\left(a,e^{2\pi i m/M}\right) e^{-2\pi i n m/M},
\end{equation}
which may be evaluated efficiently using standard FFT routines \cite{Cavers78,Marder07}. Octave/Matlab code for implementing this inversion---and hence reproducing the theory curves of Fig.~3---is available for download from  \texttt{www.ul.ie/sdcs/people/kevin-osullivan}.

\section{S3 Old-age asymptotics}\label{sec:oldage}
% see compete_paper_defence_1.pdf for notes
The large-$n$ asymptotic behaviour of $q_n(a)$ can be obtained in the limit $a\to \infty$ by asymptotic analysis of the solution of Eq.~(\ref{Gen}), with
 \begin{equation}
 \lim_{a\to\infty} H(a,x) = x\, G(\infty,x)\, f(1-\lambda+\lambda G(\infty,x)),
 \end{equation}
 where $G(\infty,x) = \lim_{a\to\infty}G(a,x)$. The following general result will prove useful (cf.~Lemma 5.3.2 of Ref.~\cite{Wilfbook}):
\begin{figure}
\centering
\epsfig{figure=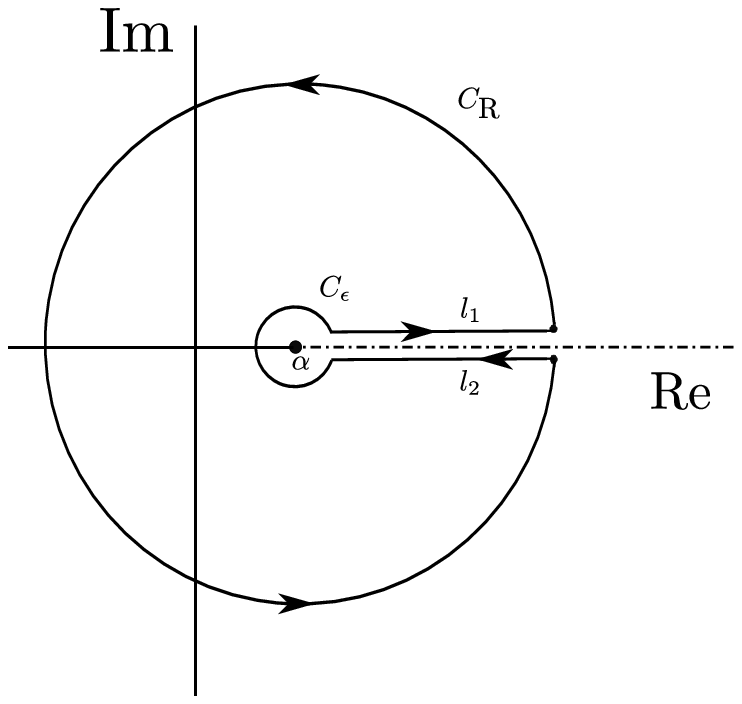,width=6.9cm}
\caption{The contour $C$ in the complex $x$-plane for the PGF inversion integral (\ref{invint}). A branch cut extends from  $\alpha$ to $\infty$ (with $\alpha=1$ for Lemma 1); the circular arcs have radii of $\epsilon$ and $R$.}\label{branch}
\end{figure}
\begin{thm}
Let $\Phi(x) = \sum_{k=0}^\infty \pi_k x^k$ be the PGF for the distribution $\pi_k$, and suppose $\Phi$ has the following asymptotic series as $x\to 1$:
\begin{equation}
\Phi(1-w) \sim \text{analytic part} + \sum_{m=1}^\infty c_m w^{\beta_m} \quad \text{ as } w\to 0,\label{PhiLemma}
\end{equation}
where $w=1-x$ and $\beta_1<\beta_2<\ldots$ are positive, non-integer powers (note that the analytic part of $\Phi$ can be written as a power series in $w$ with integer powers). Then the leading-order asymptotic behaviour of $\pi_k$ is
\begin{equation}
%\pi_k \sim \frac{c_1}{\pi}\sin(\beta_1 \pi) \Gamma(\beta_1+1) k^{-\beta_1-1} \quad \text{ as } k\to \infty \label{AsyG},
\pi_k \sim \frac{c_1 }{\Gamma(-\beta_1)} k^{-\beta_1-1} \quad \text{ as } k\to \infty \label{AsyG},
\end{equation}
where $\Gamma$ is the Gamma function.
\end{thm}
To prove this result, we begin with the inversion integral
\begin{equation}
\pi_k = \frac{1}{2 \pi i}\oint_C \Phi(x) x^{-k-1}\, dx, \label{invint}
\end{equation}
where the contour $C$ can be deformed onto the contour $C_\epsilon \cup l_1 \cup C_R \cup l_2$ shown in Fig.~\ref{branch}. The point $\alpha=1$ is a branch point, with the branch cut running from $\alpha$ to $\infty$. It is straightforward to show that the integrals along the circular contours $C_\epsilon$ and $C_R$ limit to zero as $\epsilon\to 0$ and $R\to \infty$ \cite{Keenerbook}, leaving
\begin{equation}
\pi_k = \frac{1}{2 \pi i} \left[ \int_{l_1} \Phi(x) x^{-k-1}\, dx +  \int_{l_2} \Phi(x) x^{-k-1}\, dx\right]
\end{equation}
Along the rays $l_1$ and $l_2$ we make the substitution $x=e^\rho$ to obtain integrals whose asymptotic behaviour may be determined using Watson's Lemma \cite{BenderOrszag}. The contributions from the analytic part of $\Phi$ to the $l_1$ integral and to the $l_2$ integral cancel each other. Along the branch cut, we write the leading-order non-analytic term of (\ref{PhiLemma}) as
\begin{align*}
c_1 w^{\beta_1} & = c_1 (1-x)^{\beta_1} \\
                & = c_1(1-e^\rho)^{\beta_1}\\
                & \sim c_1 (-\rho)^{\beta_1} \quad \text{ as } \rho \to 0\\
                & = \left\{ \begin{array}{ll}
                                c_1 e^{-\pi \beta_1 i } \rho^{\beta_1} & \text{ as }\rho\to 0\text{ along }l_1 \\
                                c_1 e^{\pi \beta_1 i } \rho^{\beta_1} & \text{ as }\rho\to 0\text{ along }l_2
                            \end{array}\right.
\end{align*}
to obtain
\begin{align}
\pi_k &\sim \frac{1}{2\pi i} \left[ \int_0^\infty c_1  e^{-\pi \beta_1 i} \rho^{\beta_1} e^{-k \rho} d\rho + \int_\infty^0 c_1  e^{\pi \beta_1 i} \rho^{\beta_1} e^{-k \rho} d\rho\right] \quad\text{ as } k\to \infty\nonumber\\
&= - \frac{c_1}{\pi} \sin(\pi \beta_1) \int_0^\infty \rho^{\beta_1} e^{-k \rho} d\rho.
\end{align}
The integral in this expression evaluates to $\Gamma(\beta_1+1)k^{-\beta_1-1}$. Using the Euler reflection formula $\pi/\sin(\pi\beta_1) = \Gamma(\beta_1)\Gamma(1-\beta_1)$ and the Gamma function property $\Gamma(y+1)=y\Gamma(y)$ completes the proof.

The following result allows the large-$n$ asymptotic behaviour of $q_n$ to be determined from the asymptotic form of $G(\infty,x)$:
\begin{thm}
Let $H(\infty,x) = x \, G(\infty,x) f(1-\lambda+\lambda G(\infty,x))$ be the PGF for the distribution $q_n$, and suppose
\begin{equation}
G(\infty,1-w)\sim 1-\phi(w)
\quad \text{ as } w\to 0,
\end{equation}
where $w=1-x$.  Then the asymptotic form of $H(\infty,x)$ is
\begin{equation}
H(\infty,1-w)\sim 1-w-(\lambda z+1)\phi(w)\quad \text{ as } w\to 0,
\end{equation}
and applying Lemma 1 allows the asymptotics of $q_n$ to be determined from the nonanalytic part of $\phi(w)$.
\end{thm}
The proof requires only the Taylor expansion of $f(1-\lambda+\lambda(1-\phi))$ to first order in $\phi$, noting that $\phi(0)=0$ and $f'(1)=z$.

 Turning now to the long-time (old-age) limit of Eq.~(\ref{Gen}), $G(\infty,x)$ is the solution of
 \begin{equation}
 \lambda z+\mu-(\lambda z+1)G+(1-\mu) x\, G\, f(1-\lambda+\lambda G)=0 \label{Gsteady}.
 \end{equation}
 We seek an asymptotic series solution for $G(\infty,x)$ by inserting the expression
 \begin{equation}
 G(\infty,1-w) \sim \sum_m c_m w^{\beta_m} \label{Gexpn}
 \end{equation}
 into Eq.(\ref{Gsteady}) to determine the successive values of $\beta_m$ needed to balance the leading-order powers of $w=1-x$. Then we apply Lemmas 1 and 2 to
 determine the asymptotic form of $q_n$ for large $n$ values.
\begin{itemize}
\item{\bf Case 1: $f^{\prime\prime}(1)$ infinite, $\mu=0$}\\
An out-degree distribution with power-law tail $p_k \sim D k^{-\gamma}$ with $\gamma$ between 2 and 3 has a divergent second moment, so $f^{\prime\prime}(1)$ is infinite, and the behaviour of $f(x)$ near $x=1$ is given by Lemma 1 as
\begin{equation}
f(1-w) \sim 1- z w + D \Gamma(1-\gamma) w^{\gamma-1} \quad\text{ as } w \to 0 \label{finfinite}
\end{equation}
(recall $z=f^\prime(1)$ is the mean degree). For $\mu=0$, inserting expansion (\ref{Gexpn}) into Eq.~(\ref{Gsteady}) then yields  the leading-order asymptotic behaviour of $G$ as
\begin{equation}
G(\infty,1-w) \sim 1 - \frac{1}{\lambda} \left( D \Gamma(1-\gamma)\right)^{-\frac{1}{\gamma-1}} w^\frac{1}{\gamma-1} \quad\text{ as } w \to 0,
\end{equation}
and the popularity distribution asymptotics follow from Lemmas 1 and 2:
\begin{equation}
q_n\sim B n^{-\frac{\gamma}{\gamma-1}} \quad \text{ as } n \to \infty,
\end{equation}
with prefactor
\begin{equation}
B = - (\lambda z+1)\frac{\left( D \Gamma(1-\gamma)\right)^{-\frac{1}{\gamma-1}} }{\lambda \Gamma\left(\frac{1}{1-\gamma}\right)}. \label{Beqn}
\end{equation}
 The network used in Fig.~3(b) of the main text, for example, has out-degree distribution $p_k = D k^{-\gamma}$ for $k\ge 4$, with $p_k=0$ for $k<4$: the absence of degrees less than 4 ensures that the mean degree $z$ is reasonably large ($z=10.6$ for $\gamma=2.5$), as assumed in the theory. The normalization constant $D$ for this case is
\begin{equation}
D = \frac{1}{\zeta(\gamma)-1-2^{-\gamma}-3^{-\gamma}},
\end{equation}
where $\zeta$ is the Riemann zeta function,
%and the mean degree is
%\begin{equation}
%z = \frac{\zeta(\gamma-1)-1-2^{1-\gamma}-3^{1-\gamma}}{\zeta(\gamma)-1-2^{-\gamma}-3^{-\gamma}}.
%\end{equation}
%Inserting these values of $z$ and $D$ into
allowing the prefactor $B$ to be calculated explicitly from Eq.~(\ref{Beqn}).

\item{\bf Case 2: $f^{\prime\prime}(1)$ infinite, $0<\mu<1$}\\ When  $f$ is given by Eq.~(\ref{finfinite}) and $\mu>0$, the leading-order behaviour of $G$ is found to have the form
\begin{equation}
G(\infty,1-w)\sim 1 - \frac{1-\mu}{\mu(\lambda z +1)} w +D \Gamma(1-\gamma) \lambda^{\gamma-1} \left[ \frac{1-\mu}{\mu(\lambda z+1)}\right]^\gamma w^{\gamma-1} \quad\text{ as } w\to 0
\end{equation}
and the first non-analytic term gives the asymptotic form of the popularity distribution from Lemmas 1 and 2:
\begin{equation}
q_n \sim C n^{-\gamma} \quad \text{ as } n \to \infty.
\end{equation}
The prefactor of the popularity distribution power-law is explicitly related to the prefactor of the degree distribution by
\begin{equation}
C = \frac{D(\lambda z+1)}{\lambda}\left(\frac{\lambda(1-\mu)}{\mu(\lambda z+1)}\right)^\gamma.
\end{equation}

\item{\bf Case 3: $f^{\prime\prime}(1)$ finite, $\mu=0$}\\
 If $f^{\prime\prime}(1)$ is finite (so the network out-degree distribution $p_k$ has finite second moment) and $\mu=0$, we expand $f(1-\lambda+\lambda G)$ to second-order terms in the small parameter $\phi=1-G$, and find the leading-order asymptotics of $G(\infty,1-w)$ to be
\begin{equation}
G(\infty,1-w) \sim 1-\left( \lambda z + \frac{1}{2}\lambda^2 f^{\prime\prime}(1)\right)^{-\frac{1}{2}} w^\frac{1}{2} \quad\text{ as }w\to 0.
\end{equation}
Using  Lemmas 1 and 2, we conclude that
\begin{equation}
q_n(\infty)\sim A\, n^{-\frac{3}{2}} \quad \text{ as } n\to \infty,
\end{equation}
where the prefactor is
\begin{equation}
A=(\lambda z+1)\left[ 2 \pi \lambda \left(\lambda f^{\prime\prime}(1)+2 z\right)\right]^{-\frac{1}{2}}.\label{Aeqn}
\end{equation}

\item{\bf Case 4: $f^{\prime\prime}(1)$ finite, $0<\mu\ll 1$}\\ Finally, we consider the case where $\mu>0$ but $f^{\prime\prime}(1)$ is finite. In this case, $G(\infty,x)$ is analytic at $x=1$, indicating that the popularity distribution does not have a power-law tail, i.e., Lemma 1 does not apply. However, if we write $G(\infty,1-w)=1-\phi(w)$ with $|\phi|\ll1$ and expand Eq.~(\ref{Gsteady}) for $w$ near 0, retaining terms of orders $w$, $\phi$ and $\phi^2$, but neglecting terms of order $w \phi$, we obtain a quadratic equation for $\phi$ with solution
\begin{equation}
\phi(w) = \frac{-\mu(\lambda z+1)+\sqrt{\mu^2(\lambda z +1)^2+2 \lambda(1-\mu)^2\left(\lambda f^{\prime\prime}(1)+2 z\right) w}}{(1-\mu)\lambda\left(\lambda f^{\prime\prime}(1)+2 z\right)}.\label{phi}
\end{equation}
Note that as $\mu\to 0$ this solution scales as $\phi = O\left(w^\frac{1}{2}\right)$ as $w \to 0$, which is consistent with the choice of retained terms in the scaling analysis of  Eq.~(\ref{Gsteady}). The solution (\ref{phi}) has a branch point in the complex $x$-plane at (recall $w=1-x$):
\begin{equation}
 \alpha= 1 +\frac{\mu^2(\lambda z +1 )^2}{2 \lambda(1-\mu)^2 \left(\lambda f^{\prime\prime}(1)+2 z\right)}>1.
\end{equation}
Integrating along the branch cut in the complex $x$-plane (in a very similar fashion to the proof of Lemma 1)
 % see Defence on 12.233
enables us to find the large-$n$ asymptotic form of the popularity distribution as
\begin{equation}
q_n(\infty) \sim A\, n^{-\frac{3}{2}} e^{-\frac{n}{\kappa}} \quad \text{ as } n \to \infty \label{q4}
\end{equation}
with $A$ given by Eq.~(\ref{Aeqn}) and the large-$n$ cutoff by
\begin{equation}
    \kappa=\frac{2\lambda(\lambda f^{\prime\prime}(1)+2 z)}{\mu^2(\lambda z+1)^2},
\end{equation}
and where we have assumed $\mu \ll 1$ to simplify the results. When $\lambda=1$, Eq.~(\ref{q4}) reduces to
%Eq.~(4) of the main text.
the result given in the main text.
\end{itemize}

{
\section{S4 Assumptions of the model, and relevance to real networks}

\begin{figure}
\centering
\epsfig{figure=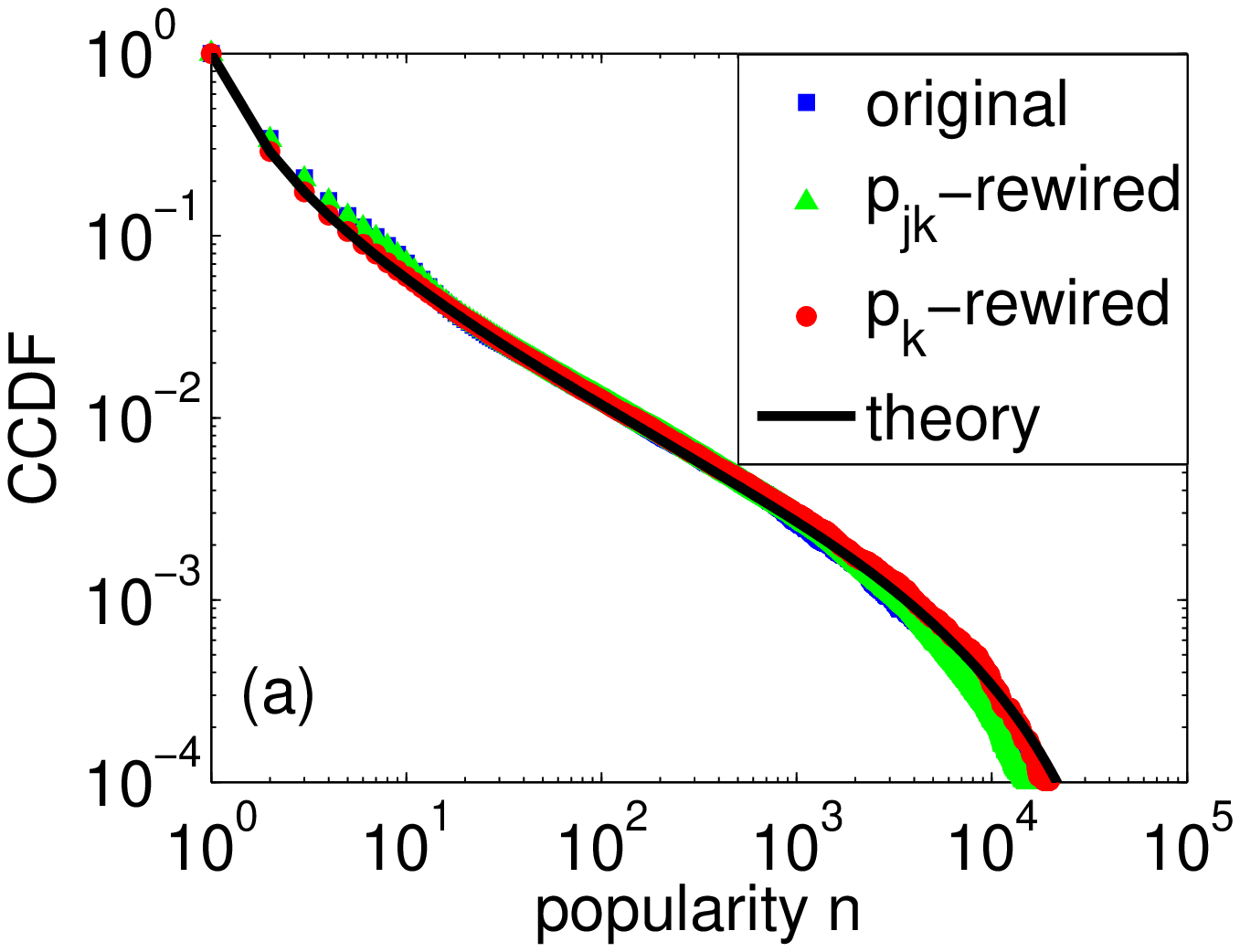,width=5.5cm}
\epsfig{figure=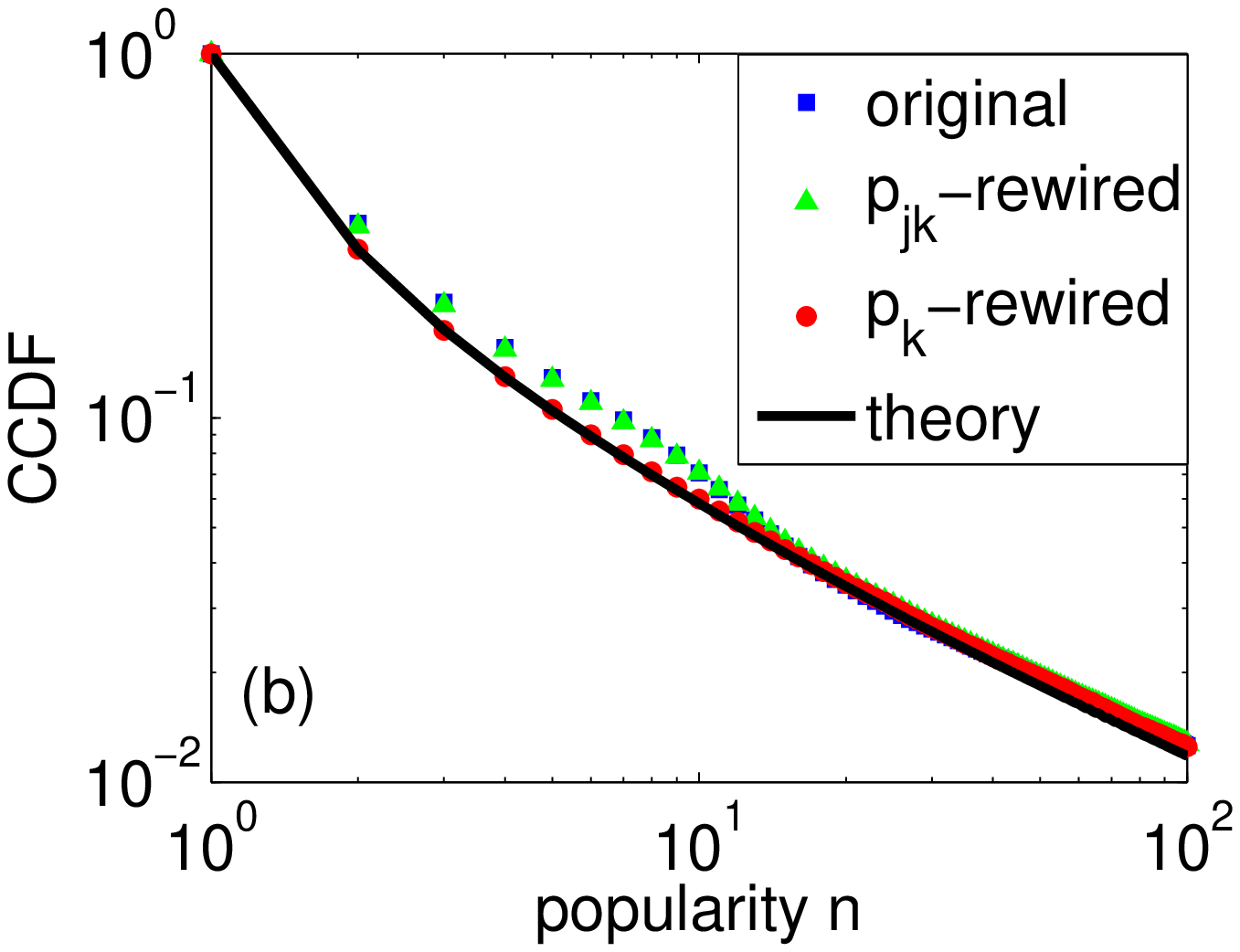,width=5.5cm}
\epsfig{figure=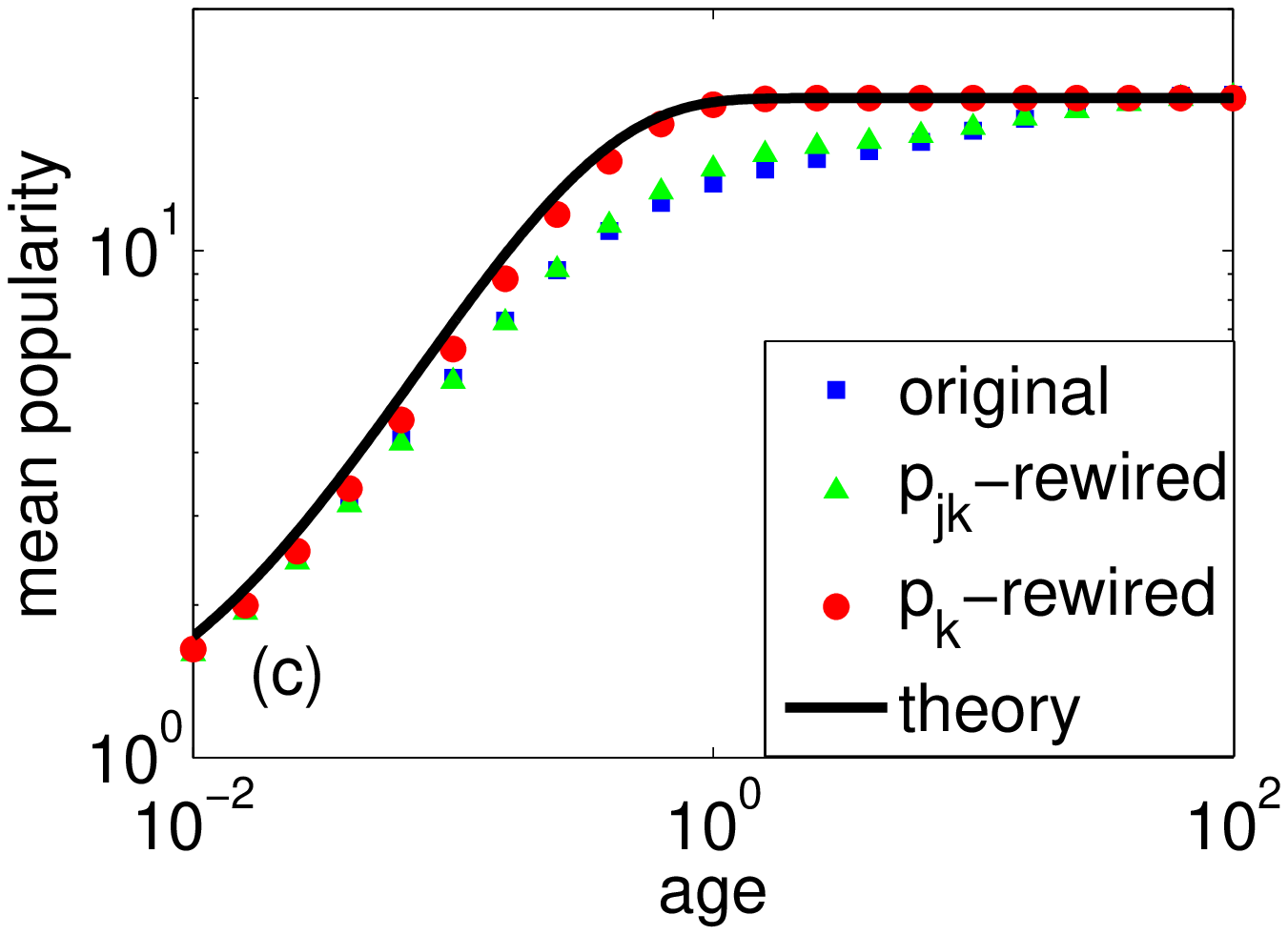,width=5.5cm}
\caption{ {  Popularity distributions at age $a=10$ ((a) and (b)) and mean popularity (c) from simulations run on the original Spanish 15M network \cite{Borge11,Gonzalez11} (as Fig.~3(c)), the $p_{jk}$-rewired network, and the $p_k$-rewired network, compared with the theory (black curves). Note the close agreement of results on the original network with the results on the $p_{jk}$-rewired network.}}\label{fignewSI}
\end{figure}

In deriving the branching-process theory for the model in Sec.~S1, we assume that the underlying network is tree-like, i.e., that no loops exist. We also assume that every node follows (approximately) $z$ other nodes, so the in-degree distribution is homogeneous. Both of these assumptions are known to be invalid for real-world networks such as Twitter; for example, in the Spanish 15M network\footnote{{The network is available from \texttt{http://cosnet.bifi.es/research-lines/online-social-systems/15m-dataset}}} used in Fig.~3(c) \cite{Borge11,Gonzalez11}, 44\% of all links are reciprocal links (i.e., node $A$ follows node $B$ and node $B$ follows node $A$), whereas a truly tree-like network would have no reciprocal links. Despite the violation of the assumptions used to derive it, we find the tree-based theory performs rather well when compared with numerical simulations of the model on real-world networks (Fig.~3(c)). This finding is consistent with the effectiveness of tree-like theory in the context of other network dynamics \cite{Melnik11,Gleeson12} and merits some further attention here.

Following \cite{Melnik11,Gleeson12}, we run simulations of the model on the original real-world network and on two rewired versions of the network.  The first rewiring is called \emph{$p_{jk}$-rewiring}, as it preserves the joint distribution $p_{jk}$ of in- and out-degrees in the network. To generate the $p_{jk}$-rewired network, we begin by severing all links of the original network, but allow each node to retain the same number $j$ of ``in-stubs'' and $k$ of ``out-stubs'' that represent the endpoints of the deleted original edges. We then randomly select one out-stub and one in-stub and create a new edge joining the selected nodes. By repeating this procedure (each time selecting from the set of unused out-stubs and in-stubs) we create a network with precisely the same $(j,k)$ distribution as the original network, but with very few short cycles or loops (i.e., clustering is reduced to the level of the corresponding directed configuration-model network \cite{Colomer13}).

The second rewiring process we consider is called \emph{$p_k$-rewiring}: it preserves the out-degree distribution $p_k$ of the original network, but replaces the in-degree distribution with a Poisson distribution. As above, we begin by severing all links of the original network, retaining the original number $k$ of out-stubs for each node.  The edges of the new network are created by randomly selecting an unused out-stub and creating a directed edge from the selected out-stub to a node chosen uniformly at random from the $N-1$ other nodes in the network. When all rewired edges are inserted, the resulting new network has the same out-degree distribution $p_k$ as the original network, but has a Poisson in-degree distribution, and also (as for the $p_{jk}$-rewired case) has substantially fewer loops that the original network (i.e. clustering is much reduced).

Figure~\ref{fignewSI}(a) shows simulation results (for the popularity distribution at age $a=10$) for the original network (blue squares), the $p_{jk}$-rewired network (green triangles), and the $p_k$-rewired network (red circles), along with the theory (black curves). Figure~\ref{fignewSI}(b) is a magnification of the low-$n$ part of Fig.~\ref{fignewSI}(a), to highlight the differences between the various cases; Fig.~\ref{fignewSI}(c) shows the mean popularity of age-$a$ memes $m(a)=\sum n q_n(a)$  for the three networks. It is clear that the theory curves match very closely to the numerical simulations on the $p_k$-rewired network; indeed, the red symbols and black curves are almost indistinguishable. This is to be expected, as the $p_k$-rewired network satisfies the assumptions underlying the derivation of our theory: the network is approximately tree-like, with a Poisson in-degree distribution (for large $z$ this is a homogeneous distribution). Focussing next on those cases where the results for the original network deviate from the theory, it is striking  that results for the $p_{jk}$-rewired network are very close to those of the original network. This implies that the differences observed between theory and simulations on the original network are due chiefly to the in-degree distribution being non-Poissonian, while the effect of clustering and loops---which are strongly present in the original network but absent in the $p_{jk}$-rewired case---appears to be almost negligible. Since the results on the original network are very similar to the results for the $p_{jk}$-rewired network, it follows that a  theory that incorporates information on the in-degree distribution of the network---but retains the tree-based derivation appropriate for zero-clustering networks---should be able to closely match to the simulation results on  the original (clustered) network. The derivation of such a theory is the focus of current work.
} % end blue section

{
\section{S5 A modified model, with restricted retweeting}
A feature of the model described in the main text is that memes remain on users' screens even after the user retweets the meme, and so the same user may retweet a single meme multiple times. This feature is intended to model memes such as hashtags that may be used in multiple message sent by a single user. However, if the memes under consideration are entire messages, then this feature of the model is unrealistic, as each Twitter message is usually retweeted at most once by any single user. In this section we therefore consider a modified version of the basic model and show that criticality is retained in the modified model.

\begin{figure}
\centering
\epsfig{figure=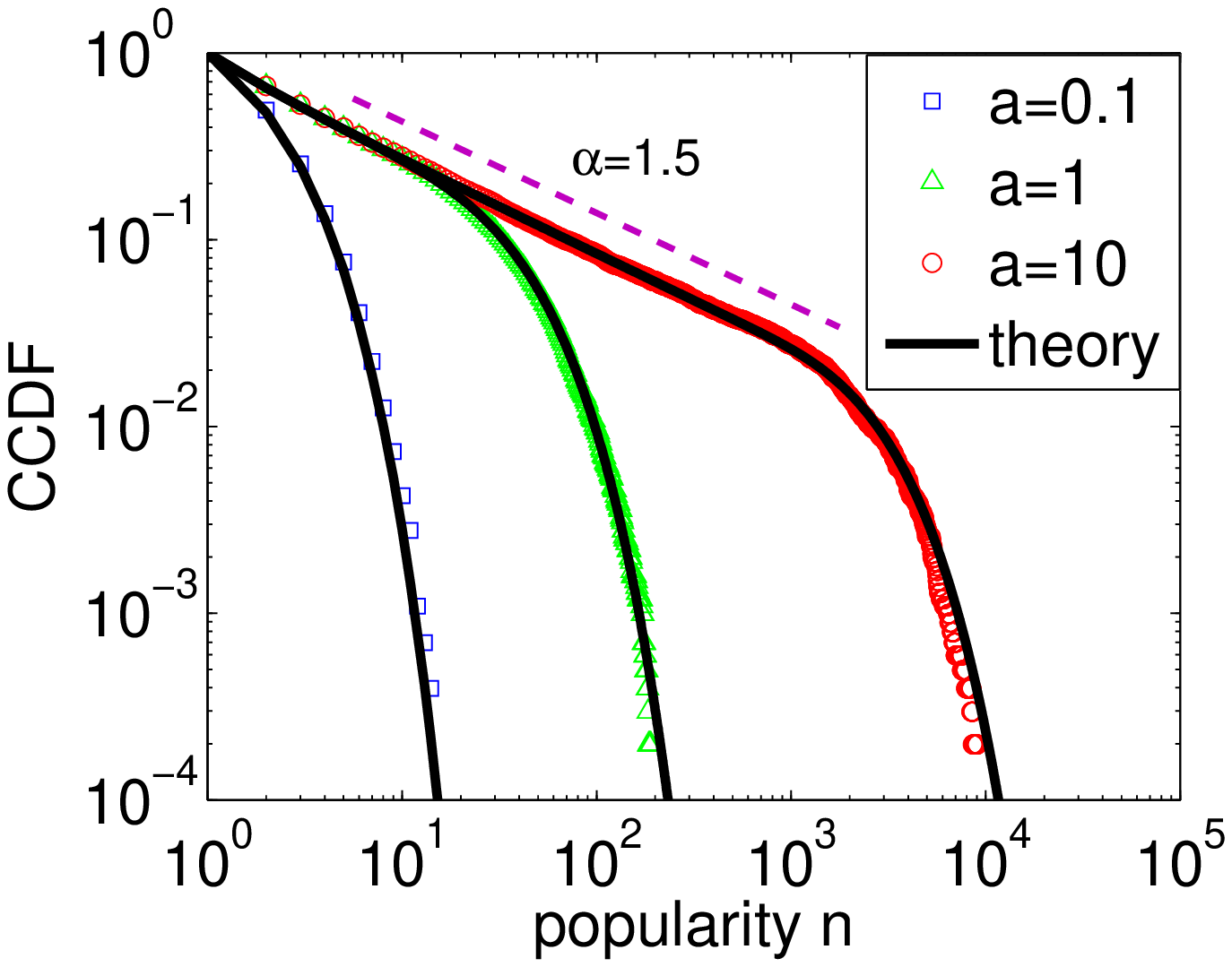,width=6.5cm}
\caption{{Numerical simulations using modified rule on retweeting, compared with the theory of Eqn.~(\ref{Gkeqn2mod}), on the same network as Fig.~3(a), with $\mu=0$.}}\label{Smod}
\end{figure}
The numerical simulation results shown in Fig.~\ref{Smod} are for the same network as used in Fig.~3(a): $z$-regular out-degrees, with $z=10$, and with $\mu=0$. However, here the model rules are altered so that each node keeps a list of the memes it has tweeted in the past. When a node is selected to retweet, it first checks whether the meme on its screen has already been retweeted by it; if so, nothing happens, but if not, retweeting occurs as usual. Figure~\ref{Smod} shows very similar results to Fig.~3(a), including a power-law popularity distribution in the infinite-age limit with exponent $\alpha=3/2$. This indicates that the criticality mechanism is not strongly affected by the modification of the model.

Furthermore, analytical insight can be gained by considering a slightly different modification of the model rules. Here we impose the new rule that after any tweeting event the screen of the tweeting node is set to the ``empty'' state. Like the first modification, this rule prevents successive tweets of a node being of the same meme (though it does allow for a meme to be ```rediscovered'' by a node some time subsequent to its first tweeting). A consequence of this modification is that the system contains a non-zero fraction of empty screens at any time $t$. Taking $\mu=0$, $\lambda=1$ and $c=1$ for simplicity of presentation, we can obtain the steady-state fraction $i$ of non-empty (or ``infected'') screens using the mean-field approach employed in the analysis of the two-meme case in the main text. When a retweet event occurs, the subsequent emptying of the tweeting node's screen reduces the number of infected screen by one, but the broadcast to (on average) $z$ followers leads to $z(1-i)$ previously empty screens becoming non-empty. The overall expected change in the number of infected screens is thus $-1+z(1-i)$, and in the dynamical equilibrium this should be zero, yielding
\begin{equation}
i=\frac{z-1}{z}.
\end{equation}

We now consider how the outcomes listed in section S1 for a particular (non-empty) screen $S_1$ are altered under this modified rule. Outcome (a)---the overwriting of screen $S_1$ by a meme tweeted by another node---now occurs with probability $z i \Delta t = (z-1)\Delta t$, with the factor of $i$ giving the probability that the tweeting node has a non-empty screen. Outcome (c) occurs with the same probability ($\Delta t$) as in the original model, but the branch on screen $S_1$ is terminated by the emptying of the screen after tweeting, giving the new overall contribution of outcome (c) as $\Delta t\, x\, \left[ G(a-\Delta t,x)\right]^k$. The probability of outcome (d) is now $1-(z\, i \,\Delta t +\Delta t) = 1-z \Delta t$, and so the expression for $G^{(k)}(a,x)$ correct to first order in $\Delta t$ (and recall $\mu=0$ here) is, cf.~Eq.~(\ref{Gkeqn0}):
\begin{equation}
%\begin{align}
%G^{(k)}(a,x) &=\nonumber \\
G^{(k)}(a,x) = \underbrace{(z-1)\, \Delta t}_{\text{(a)}}+\underbrace{\Delta t\, x\, [G(a-\Delta t,x)]^k}_\text{(c)} +\underbrace{(1-z\,\Delta t)G^{(k)}(a-\Delta t,x)}_{\text{(d)}}\label{Gkeqn0mod}.
%\end{align}
\end{equation}
In the limit $\Delta t \to 0$, we obtain the following differential equation for $G^{(k)}(a,x)$, cf.~Eq.~(\ref{Gkeqn}):
\begin{equation}
\frac{\partial G^{(k)}}{\partial a} = z-1-z\, G^{(k)}+ x\, [G]^k \label{Gkeqnmod},
\end{equation}
and multiplying by $p_k$ and summing over all $k$ gives the following equation for $G(a,x)$, cf.~Eq.~(\ref{Gkeqn2}):
\begin{equation}
\frac{\partial G}{\partial a} = z-1-z\, G+ x\, f(G) \label{Gkeqn2mod}.
\end{equation}
The popularity distributions produced by this equation are generated by the function $H(a,x)=x f(G(a,x))$  and are shown by the curves in Fig.~\ref{Smod}; note their close resemblance to the numerical simulation results, despite the latter being implemented using the more severe form of the unique-retweeting rule.

Equation (\ref{Gkeqn2mod}) can be analyzed using the same methods as for Eq.~(\ref{Gen}), and reveals the same critical behavior: the mean popularity of age-$a$ memes grows linearly with age, and the old-age asymptotics of the popularity distribution are essentially the same as the $\mu=0$ cases determined in Sec.~S3.

} % end blue section

\bibliography{../compete_bib}
\end{document}